\documentclass[letterpaper, 10 pt]{article}  

\usepackage[latin1]{inputenc}

\usepackage{tikz}
\newcommand{\co}{130}
\newcommand{\cod}{190}
\definecolor{pink}{RGB}{\co, \co, \co}
\definecolor{lpink}{RGB}{\cod, \cod, \cod}
\definecolor{blue}{RGB}{\co,\co,\co}
\definecolor{lblue}{RGB}{\cod,\cod,\cod}
\usepackage{verbatim}
\usepackage{amsmath}
\usepackage{amsthm}
\newtheorem{thm}{Theorem}
\newenvironment{thmpri}
  {%
   \begin{thm}}
  {\end{thm}}
\newenvironment{thmbis}[1]
  {%
   \addtocounter{thm}{-1}%
   \begin{thm}}
  {\end{thm}}
\newtheorem{cor}{Corollary}
\newtheorem{defi}{Definition}
\newtheorem{prop}{Proposition}
\newenvironment{proppri}
  {%
   \begin{prop}}
  {\end{prop}}
\newenvironment{propbis}[1]
  {%
   \addtocounter{prop}{-1}%
   \begin{prop}}
  {\end{prop}}
\newtheorem{lem}{Lemma}

\newtheorem{prob}{Problem}
\newtheorem{ex}{Example}
\newtheorem{ass}{Assumption}
\newtheorem{syst}{System}
\usepackage{cite}

\usepackage{color}
\usepackage{amsfonts}
\usepackage{amssymb}
\usepackage{graphicx}
\usepackage{algorithmicx}
\usepackage{algpseudocode}
\usepackage{color}
\usepackage{ulem}
\title{\LARGE \bf
Efficient Algorithms for the Consensus Decision Problem \thanks{ 
All authors are members of ICTEAM, Universit\'e catholique de Louvain, Belgium. R. M. Jungers is an F.R.S.-FNRS research associate. Their work is supported by the Belgian Network DYSCO, funded by the Belgian government and the Concerted Research Action (ARC) of the French Community
of Belgium. 
{\tt \{pierre-yves.chevalier, julien.hendrickx, raphael.jungers\}@uclouvain.be}}}

\author{P.-Y. Chevalier \and J. M. Hendrickx  \and R. M. Jungers}

\begin{document}

\maketitle
\thispagestyle{empty}
\pagestyle{empty}

\section*{Abstract} We address the problem of determining if a discrete time switched consensus system converges for any switching sequence and that of determining if it converges for at least one switching sequence.
For these two problems, we provide necessary and sufficient conditions that can be
checked in singly exponential time.

As a side result, we prove the existence of a polynomial time algorithm for the first problem when the system switches between only two subsystems whose corresponding graphs are undirected. The NP-hardness of this problem had been left open by Blondel and Olshevsky \cite{BO}. 

\section{Introduction}

The problem of how a group of agents can reach agreement on some value has attracted an important research effort. The need for coordination schemes is present in many applications such as autonomous platoons of vehicles \cite{coh}, data fusion in systems with distributed measurements \cite{olf, sensor2}, distributed optimization \cite{ned} or coordination of groups of mobile agents (see \cite{tac} and references therein).  \emph{Consensus systems} describe the dynamics of these coordination procedures. They have also been used as models for natural phenomena such as flocking \cite{Vicsek} or opinion dynamics \cite{Krause}. See also \cite{surv, Mesbahi} for a survey.

In many of these systems, the agents update their value by taking a weighted average of the values of agents with which they can communicate: 
\begin{align} & x_i(t) = \sum_j a_{ij}(t) x_j(t-1) \label{update} \\ 
& \text{with } a_{ij}(t) \geq 0 \text{ and } \sum_j a_{ij}(t) = 1 \label{stoch} \end{align} where $x_i$ is the value of agent $i$ and $a_{ij}$ represents the way agent $j$ influences agent $i$. 
Matrices whose elements satisfy (\ref{stoch}) are called \emph{stochastic}.
Agents following these dynamics tend to be more and more in agreement, in the sense that their values generally get closer to each other. A general question is to know if the system converges to a state of consensus, i.e., a state in which all agents have the same value. 
In a large class of systems, the interaction coefficients $a_{ij}(t)$ depend on the state $x$, making the system nonlinear \cite{tac, Krause}. Deciding whether the system converges to consensus is therefore a hard problem. For example, \cite{2R} presents a relatively simple model for which no conditions for convergence to consensus are known.

In some situations, even if it is hard to explicit the complete  sequence of matrices $A(t)$ corresponding to System (\ref{update}), it may be possible to guarantee that these matrices remain in some set $S$. 
In this article, we study convergence conditions based only on the knowledge of the set $S$.

\vspace{.5cm}

Blondel and Olshevsky studied the complexity of deciding if for a given set $S$, all trajectories converge to consensus \cite{BO}. They based some of their results on work by Paz on heterogeneous Markov chains which also involve long products of stochastic matrices \cite{Paz}. They proved that the problem is decidable and that it is NP-hard for sets of at least two matrices. 
From the decidability proof, a doubly exponential decision algorithm can be deduced. 
They raise the question of the existence of a singly exponential decision algorithm. 
 When restricting the problem to undirected graphs of communication, they proved NP-hardness for sets of at least three matrices while the case of sets of two matrices is left open.

\vspace{.5cm}

Many consensus systems are in fact switched systems (for example \cite{optimal_switch, CDC}) for which convergence questions have been extensively covered in the literature (see for example \cite{book, Lib, S10}). 
In particular, geometric techniques have been developed to prove convergence, some of them using the existence of invariant sets \cite{Bara, variat}.
The link between consensus and the asymptotic stability of switched systems has been mentioned in \cite{tac} and \cite{CDC}. In \cite{MTNS}, we explored this link as we have proven that the decidability result of \cite{BO} can be retrieved using techniques from switched systems theory. 
In this article, we push further this approach of using techniques inspired by switched systems theory.

\vspace{.5cm}

In this article, using ideas from the theory of switched systems, we obtain new conditions for convergence in the general case and we prove that these conditions can be checked by a singly exponential algorithm.  
We also show that the case of sets of two matrices with undirected communication graph can be solved in polynomial time.
We moreover consider the problem of the existence of a converging trajectory.
Our theorems are valid under an assumption that is slightly weaker than requiring the matrices to be stochastic.  

\subsection{Outline}

The next section is dedicated to the formulation of the problem. 
In Section \ref{inv}, we show that our system admits an invariant polyhedron. In Section \ref{Finiteness}, we prove that the convergence of the system can be predicted by looking only to trajectories up to a finite time. In Section \ref{graph}, we represent trajectories as paths on a graph.  
We prove that deciding Problems 1 and 2 are then equivalent to simple decision problems on this graph. 
In Section \ref{Complexity}, we prove that the complexity of deciding Problems 1 and 2 is singly exponential.
In the last section, we focus on sets of two undirected matrices (see Section \ref{new_part} for a precise statement).
We show the existence of a polynomial-time algorithm for this case.

\section{Problem Setting}

Let  $S = \{A_1, \dots , A_m\}$ be a set of matrices that share a common eigenvector $\mathbf{1} = \begin{pmatrix} 1 & \dots & 1 \end{pmatrix}^\top$ of eigenvalue 1. We study the following system
\begin{syst}
\begin{equation*} 
\begin{aligned} x(t+1) &= A_{\sigma(t)} x(t) \\
x(0) &= x_0
\end{aligned} 
\end{equation*}
\label{sys}
\end{syst}
where $\sigma: \mathbb{N} \mapsto \{1, \dots, m\} : t \mapsto \sigma(t)$ is an infinite sequence of indices. Let $\Sigma$ denote the set of such sequences and $\Sigma_t$ the set of sequences limited to length $t$. We call a consensus state any multiple of $\mathbf{1} .$ We call the \emph{trajectory} of the system the sequence of iterates generated by $x_0, \sigma$.

To represent the distance to consensus, we use the following seminorm $$\|x\|_\mathcal{P} = \frac{1}{2}(\max_i{x_i} - \min_i{x_i}).$$ 

In this article, we restrict our attention to sets $S$ for which this seminorm is a common Lyapunov function in a weak sense, by which we mean that, $S$ satisfies the following assumption. 
\begin{ass} For any matrix $A \in S$, \begin{equation}\|Ax\|_\mathcal{P} \leq \|x\|_\mathcal{P}.\label{ass_single} \end{equation}
\end{ass}
Sometimes we also say that a single matrix satisfies the assumption if it satisfies inequality (\ref{ass_single}).  

Many results on consensus rely on the nonnegativity of the matrices. This nonnegativity is equivalent to the monotonicity of the system (see \cite{MDS} to see how monotonicity can help to prove convergence). We stress that our approach does not assume nonnegativity. In particular, Assumption 1 is weaker than assuming that the matrices are stochastic\footnote{Nonnegative matrices satisfying $A \mathbf{1} = \mathbf{1}$}. This assumption is made in \cite{BO} and is common for linear discrete time consensus systems.

We study two decisions problems on System \ref{sys}:
\begin{prob}[Asymptotic stability]
Given a set of matrices that share a common eigenvector $v$ of eigenvalue 1, does System \ref{sys} converge to a multiple of $\mathbf{1}$ for any initial condition $x_0$ and any sequence $\sigma$? 
\label{prob1}
\end{prob}

\begin{prob}[Reachability of consensus]
Given a set of matrices that share a common eigenvector $v$ of eigenvalue 1, is there a sequence $\sigma$ such that, for any initial condition $x_0$, System \ref{sys} converges to a multiple of $\mathbf{1}$ ?
\label{prob2}
\end{prob}

\section{Invariant polyhedron}
\label{inv}

\subsection{Polyhedra and faces}\label{pol_f}

In this section, we show that the system admits an invariant polyhedron and we define some notions related to polyhedra. 

We call a \emph{polyhedron} a subset $\mathcal{Q}$ of $\mathbb{R}^n$  that is the intersection of a finite number of halfspaces or equivalently that can be defined by $$\mathcal{Q} = \{x \; | \; Ax \leq b\}.$$ Note that a polyhedron is not necessarily bounded. We say that a set $E$ is \emph{invariant}  with respect to a matrix $A$ if $$ AE \subseteq E.$$

\begin{defi}[Faces of a polyhedron] A non-empty subset $F$ of a polyhedron $\mathcal{Q}$ is called a \emph{face} if $F = \mathcal{Q}$ or if it can be represented as $F = \mathcal{Q} \cap \left\{x \; | \; b^\top x = c \right\}$
where $b \in \mathbb{R}^{n}$, $c \in \mathbb{R}$ are such that $$\forall x \in \mathcal{Q}, \; b^\top x \leq c.$$ If the face contains $n+1$ affinely independent points, we call $n$ the \emph{dimension} of the face. 
We call a proper face a face that is not equal to $Q$.
\end{defi}

For example, the faces of a square are the square itself, the four corners and the four sides.

We call an \emph{open face} the relative interior of a face. In particular, if the face is a single point, the corresponding open face is the face itself.

A face of dimension $n-1$ is called a \emph{facet}. For a facet, there is a unique 
hyperplane $b^\top x = c$ such that the facet is equal to $ \mathcal{Q} \cap \left\{x \; | \; b^\top x = c \right\}$. We call  $b^\top x \leq c$ the \emph{facet constraint}. We say that the constraint is \emph{active} at a point $x$ when $b^\top x = c$.

The next lemma shows how the facet inequalities define the polyhedron. 

\begin{lem}[Theorem 8.1 in \cite{Faces}] Let $\mathcal{Q} = \{x \; | \; Ax \leq b\}$ be a polyhedron and let $Ax \leq b$ be non redundant constraints (no row of $Ax \leq b$ can be removed without changing $ \{x \; | \; Ax \leq b\}$). A subset $F$ of $\mathcal{Q}$ is a facet if and only if $$F = \{x \in \mathcal{Q} \; | \; a_i^\top x = b_i\}$$ for $a_i^\top x = b_i$ a row of $Ax = b$. 
\label{one_to_one}
\end{lem}

We now present a lemma that allows to represent the open faces in terms of the inequalities that define the polyhedron. 

\begin{lem} Let $\mathcal{Q} = \{x \; | \; Ax \leq b\}$ be a polyhedron. A non-empty subset $F$ of $\mathcal{Q}$ is an open face of $\mathcal{Q}$ if and only if it can be written as $$\{x \; | \; A'x = b', \; A''x < b'' \},$$ where $A'x = b'$ is a subset of the rows of $Ax = b$ and $A''x = b''$ are the remaining rows. 
\begin{proof}
A subset of $\mathcal{Q}$ is a closed face if and only if it can be written as \begin{equation} \{x \in \mathcal{Q} \; | \; A'x = b'\}\label{face_eq} \end{equation} (see Section 8.3 in \cite{Faces}).
Equation (\ref{face_eq}) can be rewritten $$F =  \{x  \; | \; Ax \leq b, \; A'x = b'\},$$ making clear that a point $x \in F$ is in the relative boundary of $F$ if and only if it satisfies $A''x = b''$ for $A''x = b''$ a subsystem of $Ax = b$ that is linearly independent from $A'x = b'$. Removing this boundary yields the result.
\end{proof}
\label{rank}
\end{lem}

The combination of these two lemmas has interesting consequences. First, two different open faces differ in at least one facet constraint. That is, there is a facet constraint $a_i^\top x \leq b_i$ such that points of one of the faces satisfy $ a_i^\top x = b_i$ and points of the other satisfy $ a_i^\top x < b_i$.

The second consequence is that a polyhedron $\mathcal{Q} = \{Ax \leq b\}$ decomposes into the disjoint union of its open faces: a point $x \in \mathcal{Q}$ is in exactly one 
open face. This face is given by $\{y \; | \; A'y = b', \; A''y < b'' \}$ where $A', b'$ is the largest subsystem of $A,b$ such that $A'x = b'$ and $A'', b''$ are the remaining rows.  

\subsection{Common invariant polyhedron for System \ref{sys}}

In this section, we identify a common invariant polyhedron for all matrices satisfying Assumption 1. We characterize its faces and we count them. This characterization will allow us to represent trajectories as sequences of faces in which the state vector is (in Section \ref{graph}). 
 The number of faces will directly influence the complexity of our convergence checking algorithm.

\begin{defi}[Polyhedron $\mathcal{P}$] Let \begin{equation}\mathcal{P} = \left\{x \; | \; \frac{1}{2}(\max_i{x_i} - \min_i{x_i}) \leq 1 \right\}.\end{equation} \label{p_def} \end{defi} 
It is a polyhedron since the constraint with the max can be decomposed into a set of simple linear constraints: \begin{equation} \mathcal{P} = \bigcap_{ij} \left\{x \; | \; \frac{1}{2} (x_i - x_j) \leq 1\right\}.\label{decompose} \end{equation}
We can directly verify from this definition that $\mathcal{P}$ is invariant for any matrix that satisfies Assumption 1. 
  This means that under Assumption 1, when the state vector $x(T)$ is in $\mathcal{P}$ then $x(t)$ stays in $\mathcal{P}$ for any $t\geq T$. We notice that $\text{int}(\mathcal{P})$ is also invariant. 
We use the letter $\mathcal{Q}$ when referring to any polyhedron and $\mathcal{P}$ for this particular polyhedron. 

\vspace{.5cm}

Thanks to Lemma \ref{one_to_one}, we know that the facets of the invariant polyhedron $\mathcal{P}$ are the sets $$\mathcal{P} \cap \left\{x \; | \; \frac{1}{2} (x_i - x_j) = 1\right\}$$ for $i \neq j$ while for the faces in general, we have the next lemma.

\begin{lem} Let us call $V = \{-1, 0, 1\}^n \backslash (\{-1, 0\}^n \cup \{0, 1\}^n)$ the set of signed binary vectors that contain at least one $1$ and one $-1$ and $\mathcal{P}$ defined as in Definition \ref{p_def}. 
In each proper open face of $\mathcal{P}$, there is exactly one  element of $V$. 

From a point $x \in \partial\mathcal{P} = \mathcal{P} \backslash \textnormal{int}(\mathcal{P})$, the element of $V$ that is in the same face is  
$v = \textnormal{round}(x + (1 - \max_i x_i ) \mathbf{1})$ where $\textnormal{round}(.)$ rounds to zero any component that is not equal to $1$ or $-1$.
\begin{proof}
\textit{Existence:} 
Let us take $x$ in a given proper open face. With $e^i$ denoting the $i$th vector of the canonical basis,  the facet constraints become  $$\frac{1}{2} (e^i - e^j)^\top x \leq 1.$$ Because $$\frac{1}{2} (e^i - e^j)^\top \mathbf{1} = 0,$$ we have that $x + (1- \max_i x_i) \mathbf{1}$ satisfies a facet constraint strictly if and only if $x$ satisfies it strictly. Therefore $x + (1- \max_i x_i) \mathbf{1}$ and $x$ are in the same open face. Now since  $x \in \partial\mathcal{P} $, we have $$\max_i x_i - \min_i x_i = 2$$ and therefore $$\forall j, \; -1 \leq x_j + (1- \max_i x_i) \mathbf{1} \leq 1$$ and at least one component of $x_j + (1- \max_i x_i) \mathbf{1}$ is equal to $1$ and at least one to $-1$. We finally note that rounding to zero the elements of $x_j + (1- \max_i x_i) \mathbf{1}$ that are not equal to $1$ or $-1$ doesn't activate or deactivate any facet constraint. Therefore, $\textnormal{round}(x + (1 - \max_i x_i ) \mathbf{1})$ is an element of $V$ that is in the same face as $x$.

\textit{Unicity:} Let $v, w \in V$ and $v \neq w$. Then, there is $i$ such that $v_i \neq w_i$. Suppose $$v_i = 1 \text{ and }w_i \neq 1.$$ The others cases: $v_i = -1, \; w_i \neq -1$ and interchanging $v$ and $w$ are similar. Because $v \in V$, there is $j$ such that $v_j = -1$. We obtain $$\frac{1}{2}(e^i - e^j)^\top v = 1$$ and $$\frac{1}{2}(e^i - e^j)^\top w < 1,$$ proving that $v$ and $w$ differ in at least one facet constraint and thus are not in the same open face. 
\end{proof}
\label{v_repr}
\end{lem} 

\begin{cor}
The number of faces of $\mathcal{P}$ is $3^n -2^{n+1} + 2$.  
\begin{proof}
There is one proper face for each element of $V$ plus the non-proper face $\text{int}(\mathcal{P})$ and there are $3^n -2^{n+1} + 1$ elements of $V$. 
\end{proof}
\end{cor}

\begin{ex}
When $n = 2$, $\mathcal{P} = \left\{x \; | \; \frac{1}{2} |x_1 - x_2| \leq 1   \right\}$. The faces are
\renewcommand\labelitemi{\tiny$\bullet$} 
\begin{itemize}
\item $\{x \; | \; \frac{1}{2} (x_1 - x_2) = 1 \}$, 
\item $\{x \; | \; \frac{1}{2} (x_2 - x_1) = 1\}$,
\item $\textnormal{int}(\mathcal{P})$. 
\end{itemize}
There are thus three faces as predicted by the corollary: $3^2 - 2^3 + 2 = 3$.
\end{ex}

From (\ref{decompose}), we can see that the polyhedron $\mathcal{P}$ is symmetric around the origin: $$\mathcal{P} = - \mathcal{P}.$$ Therefore, $\text{int}(\mathcal{P}) = - \text{int}(\mathcal{P})$ and for a proper open face $F$, $-F$ is also a face and a different one. We note $\pm F$ to denote $F \cup -F$. We note $N = \frac{1}{2}(3^n -2^{n+1} + 1)$ the number of pairs of opposite proper faces. 

\section{Finiteness}
\label{Finiteness}

In this section, we prove that the convergence of the System \ref{sys} can be analysed by looking only at finite products (up to length $N$) of the transition matrices. This result is similar in spirit with \cite{BO, MTNS}.

First, we present a lemma that plays a key role in the proof of the finiteness result. It shows how all the points in a face generate similar trajectories.  
 It is similar to a claim in the proof of Theorem 4.1 in \cite{LW95}; we state it here as an independent lemma because our hypothesis are slightly different.

\begin{lem}[Lagarias and Wang] Let $S$ be a finite set of matrices having a common invariant polyhedron $\mathcal{Q}$. 
Then, for any $A \in S$ and any open face $F$ of $\mathcal{Q}   $, 
 there exists exactly one open face $G$ (possibly $\text{int}(\mathcal{Q})$) such that $$A F \; \subseteq \; G.$$
\begin{proof}  
 
We first prove by contradiction that the image $A F$ intersects at most one open face. Suppose that there were points $x_1, x_2 \in  F$ such that $Ax_1$ and $ Ax_2$ were in different open faces. These open faces differ in at least one facet constraint, with one having $b^\top x = c$ and the other $b^\top x < c$ (Lemmas \ref{one_to_one} and \ref{rank}). Without loss of generality, suppose that $b^\top A x_1 = c$ and $b^\top A x_2 < c$. 
Since $F$ is relatively open, there exists $\varepsilon > 0$ with 
$$(1-\lambda) x_1 + \lambda x_2 \in F \text{ for }-\varepsilon \leq \lambda \leq 1 + \varepsilon.$$ In particular $y = (1+\varepsilon)  x_1 + -\varepsilon x_2 \in F$ and $b^\top Ay > c$, which implies $Ay \notin \mathcal{Q}$, contradiction with $A (\mathcal{Q}) \subseteq \mathcal{Q}$. 

Now, since $\mathcal{Q}$ is equal to the disjoint union of its open faces (see Section \ref{pol_f}), all points of $AF \subset \mathcal{Q}$ belong to the same face $G$.
\end{proof}
\label{key}
\end{lem}

We now prove the finiteness result. We remind that $N = \frac{1}{2} (3^n - 2^{n+1} + 1)$ is the number of pairs of proper faces.

\begin{proppri}[Finiteness] Let $S = \{A_1, \dots, A_m\}$ be a set of matrices satisfying $A\mathbf{1} = \mathbf{1}$ and $\|Ax\|_\mathcal{P} \leq \|x\|_\mathcal{P}$ (Assumption 1). The answer to Problem 1 (Asymptotic stability) is negative if and only if  \begin{equation} \exists \; \textnormal{proper face } F , \; \sigma \in \Sigma, \; k \leq N \textnormal{ s.t. } A_{\sigma(k-1)} \dots A_{\sigma(0)}F \subseteq \pm F. \label{cond} \end{equation} 
\begin{proof}
\textit{If.} If 
Condition (\ref{cond}) is satisfied,  taking an  initial condition in $ F$ and the sequence $$ \sigma(0),  \dots, \sigma(k-1),\sigma(0),  \dots, \sigma(k-1), \dots  $$ yields a non-converging trajectory: $$\forall T, \; x(2kT) \in F$$
and $F$ is at positive distance to consensus.

\noindent \textit{Only if.}  
For the necessity, we will prove that if Condition (\ref{cond}) is  not satisfied then $$\forall x_0, \sigma , \; \|x_0\|_\mathcal{P} = 1 \; \Rightarrow \;  \|x(N)\|_\mathcal{P} \leq r < 1  $$
for some $r$.  
 In turn, $x(t)$ will be become arbitrary close to $\text{span}\{\mathbf{1}\}$ (the subspace in which agents are at consensus). We finally prove that $x(t)$ has a limit in that subspace.

Let us fix $x_0$ and $\sigma$.
The set $X = \{x(0), \dots, x(N)\}$ contains $N+1$ elements.
Recall that by Lemma \ref{key}, if $x(i), x(j) \in \pm F$ and $0 \leq i < j \leq N$, then $$A_{\sigma(j-1)} \dots A_{\sigma(i)}F \subseteq \pm F.$$
Therefore, if Condition (\ref{cond}) is not satisfied, then there is no face $F \in \mathcal{F}$ 
such that $\pm F$ contains two elements of $X$. Since there are only $N$ pairs of opposite faces, and $N+1$ elements in $X$, 
 we conclude that there is $i$ such that $x(i) \in  \text{int}(\mathcal{P})$ and because $ \text{int}(\mathcal{P})$ is invariant, then $$x(N) \in  \text{int}(\mathcal{P}).$$ 
Because this is true for all $x_0, \sigma$, we have:
\begin{equation} \forall x_0, \sigma , \; \|x_0\|_\mathcal{P} = 1 \; \Rightarrow \;  \|x(N)\|_\mathcal{P} < 1 \label{N_int}. \end{equation}

We now prove the slightly stronger statement 
\begin{equation} \forall x_0, \sigma , \; \|x_0\|_\mathcal{P} = 1 \; \Rightarrow \;  \|x(N)\|_\mathcal{P} \leq r < 1  \end{equation}
where $$r  = \sup_{x_0 \in \partial \mathcal{P}  , \; \sigma \in \Sigma_N} \|x(N)\|_\mathcal{P}.$$ We need to prove $r < 1$ (the other inequality follows from the definition of the $\sup$).
Because $\|.\|_\mathcal{P}$ is invariant in the direction of $\mathbf{1}$ and because $\mathbf{1}$ is an eigenvector of eigenvalue 1, the value of the supremum does not change if $x_0$ is restricted to belong to $ \{x \; | \; \mathbf{1}^\top x = 0\}$:
\begin{equation} r  =  \sup_{x_0 \in \partial \mathcal{P} \cap \{x \; | \; \mathbf{1}^\top x = 0\} , \sigma \in \Sigma_N} \|x(N)\|_\mathcal{P}
.\end{equation} 
By compactness of $\partial \mathcal{P} \cap \{x \; | \; \mathbf{1}^\top x = 0\}$ and $\Sigma_N$, we obtain that the supremum  is attained:
$$r = \max_{x_0 \in \partial \mathcal{P} \cap \{x \; | \; \mathbf{1}^\top x = 0\}, \; \mu} \|x(N)\|_\mathcal{P}$$
 and by (\ref{N_int}) this quantity is smaller than 1.
 
 Now, we can use the linearity of the system to show that the $\|x\|_\mathcal{P}$ decreases by a factor $r$ every $N$ steps: at step $t$, if the system is not a consensus, $\|x(t)\|_\mathcal{P} \neq 0$ and the polyhedron can be scaled such that $x$ is on the boundary: $$x(t) \in \|x(t)\|_\mathcal{P} \partial \mathcal{P}$$ and therefore
 $$x(t+N) \in r \|x(t)\|_\mathcal{P} \mathcal{P}.$$
This, combined with the fact that $\mathcal{P}$ is invariant, implies
\begin{equation}\forall t , \; \|x(t)\|_\mathcal{P} \leq r^{\frac{t}{N} - 1} \|x_0\|_\mathcal{P} \label{xP} \end{equation} 
 which also holds for $\|x(t)\|_\mathcal{P} = 0$.
 
We now prove that the limit $\lim_{t \rightarrow \infty} x(t)$ exists. 

We start by bounding the difference between two successive iterates. Using $P = I - \frac{\mathbf{1}\mathbf{1}^\top}{n}$ and  $A \mathbf{1} = \mathbf{1}, \; \forall A \in S$, we obtain
\begin{equation} \begin{aligned} x(t+1) - x(t) &= (A_{\sigma(t)}-I) x(t) \\
& =  (A_{\sigma(t)}-I) Px(t) .
\label{x-x}
\end{aligned} \end{equation}
Using (\ref{xP}), (\ref{x-x}) and $\|Px(t)\|_\infty \leq 2 \|Px(t)\|_\mathcal{P} = 2 \|x(t)\|_\mathcal{P}$, we obtain, for some appropriate constant $C$,

\begin{equation*}
\begin{aligned}
\|x(t+1) - x(t)\|_\infty & \leq \|A_{\sigma(t)} - I \|_\infty \|Px(t)\|_\infty \\
& \leq (1+\max_{A_i \in S}{\|A_i\|_\infty}) 2 \|x(t)\|_\mathcal{P} \\
& \leq C  \|x(t)\|_\mathcal{P} \\
& \leq C r^{\frac{t}{N} - 1} \|x_0\|_\mathcal{P}.
\end{aligned}
\end{equation*}

Therefore, for any $q > s $ 
\begin{align*} \|x(q) - x(s)\|_\infty &= \sum_{t = s}^{q-1}\|x(t+1) - x(t)\|_\infty \\
& \leq \sum_{t = s}^{q-1} C r^{\frac{t}{N} - 1} \|x_0\|_\mathcal{P}\\
& \leq C r^{\frac{s}{N} - 1} \|x_0\|_\mathcal{P} \sum_{t = s}^{\infty}r^{\frac{t-s}{N}} \\
& \leq C_2 r^{\frac{s}{N}} \|x_0\|_\mathcal{P}
\end{align*} where $C_2$ does not depend on $s$.
Therefore the trajectory is a Cauchy sequence and converges. By (\ref{xP}) it can only converge to consensus.
\end{proof}
\label{nec_suf}
\end{proppri}

\begin{propbis}{nec_suf}
 Let $S$ be a set of matrices satisfying $A\mathbf{1} = \mathbf{1}$ and Assumption 1. The answer to Problem 2 (reachability of consensus) is
positive if and only if from any initial condition, the interior of $\mathcal{P}$ can be reached in $N$ steps :
\begin{equation}\forall x_0 \in  \mathcal{P}, \; \exists \sigma \in \Sigma, \; A_{\sigma(N-1)} \dots A_{\sigma(0)} x_0 \in \textnormal{int}(\mathcal{P}).\label{int_pb2}\end{equation}
\begin{proof}The proof is very similar to that of Proposition \ref{nec_suf}.
For the necessity, notice that if (\ref{int_pb2}) is not satisfied, then there is an initial condition in $\partial \mathcal{P}$ from which it is then impossible to reach $\text{int}(\mathcal{P})$ in $N$ steps. By Lemma \ref{key}, and because the number of pairs of opposite faces is $N$, it is impossible to reach $\text{int}(\mathcal{P})$ (and therefore the origin) from this initial condition. 

For the sufficiency, we start with this claim.

\noindent\textit{Claim.} There is a product of length at most  $N^2$ that maps every face into $\text{int}(\mathcal{P})$. We prove this claim constructively. 
Let $F_1, \dots, F_N$ be the pairs of proper faces. By hypothesis, there is a product $P_1$ of length at most $N$ such that $P_1 F_1 \subseteq \text{int}(\mathcal{P})$. For any $x \in \mathcal{P}$, 
\begin{equation}P_1 x \in P_1 F_2 \cup P_1 F_3 \cup \dots \cup P_1 F_N \cup \text{int}(\mathcal{P}) \label{union}.\end{equation}
By Lemma \ref{key}, $P_1 F_2$ is a subset of a pair of proper faces and therefore, there is a product $P_2$ of length at most $N$ such that $P_2 P_1 F_2 \subseteq \text{int}(\mathcal{P})$. Note that $P_1 F_2$ can be a subset of $F_1$ and therefore $P_1$ and $P_2$ could be equal. Because $P_1 F_1 \subseteq \text{int}(\mathcal{P})$, we have also
 $$P_2 P_1 F_1  \subseteq \text{int}(\mathcal{P}).$$ 
Therefore, for any $x \in \mathcal{P}$, 
$$P_2 P_1 x \in P_2 P_1 F_3 \cup P_2 P_1 F_4 \cup \dots \cup P_2 P_1 F_N \cup \text{int}(\mathcal{P}) .$$ 
Continuing this procedure yields the product of the claim.

The rest of the proof is the same as the proof of the above proposition
with 
$$r = \sup_{x_0 \in \partial \mathcal{P}}  \min_{\sigma \in \Sigma_{N^2}} \|x(N^2)\|_\mathcal{P} < 1.$$
We obtain that there is $\sigma$ such that 
$$\|x(t)\|_\mathcal{P}  \leq \frac{1}{r}\left (\sqrt[{N^2}]{r}\right)^\top \|x_0\|_\mathcal{P},$$
\end{proof}
\label{nec_suf2}
\end{propbis}

\section{Graph representation of the trajectories}
\label{graph}

We now present a method to represent trajectories as paths on a graph. The nodes represent faces and the edges represent the possibility to jump from one face to another using of the transition matrices.
We will show that this graph captures enough information to decide convergence to consensus.

\begin{defi}[Graph of faces] Given a finite set $S$ of matrices and  $\mathcal{Q}$ an invariant polyhedron that is symmetric around the origin ($\mathcal{Q} = - \mathcal{Q}$), we call the \textit{graph of faces} $\mathcal{G}$ the graph having 
\begin{itemize} 
\item one vertex for each pair of opposite faces of  $\mathcal{Q}$, one node representing $\text{int}(\mathcal{Q})$ that we call "node 1" by convention.
\item one edge from node $i$ to node $j$ if they correspond to faces $F_i$ and $F_j$ and there is $A \in S$ such that $A F_i \subseteq \pm F_j$. In particular there is one edge from node $i$ to node 1 if node $i$ corresponds to a face $F_i$ and there is $A \in S$ such that $A F_i \subseteq \text{int}(\mathcal{Q})$ and one edge going from node 1 to itself.
\end{itemize}
\label{def_gr}  
\end{defi}


\begin{ex} We construct the graph of faces of the set $$\left\{A = \begin{pmatrix} 0 & \frac{1}{2} \\ -1 & -\frac{1}{2} \end{pmatrix}, B = \begin{pmatrix}  -\frac{1}{4} & \frac{3}{4} \\-\frac{3}{4} &\frac{1}{4} \end{pmatrix}\right\}$$ and the polyhedron $$\mathcal{Q} = \left\{x \; | \; \|x\|_1 \leq 1\right\}.$$ 
The polyhedron is invariant for matrices $A$ and $B$ as depicted on Figure \ref{f1}.

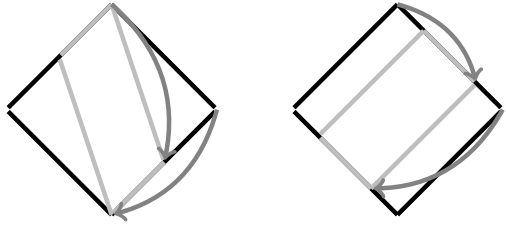
\begin{figure}[h!]
\centering

\usetikzlibrary{arrows}
\newcommand{\ec}{3.8}
\newcommand{\ba}{.7}
\newcommand{\wi}{.6}
\begin{tikzpicture}[->,shorten >=1pt,auto,node distance=2cm,
  thick,main node/.style={circle,fill=blue!20,draw,font=\sffamily\Large\bfseries}]

  \coordinate (1) {};
  \coordinate (2) at (\ba,-\ba) {};
  \coordinate (3) at (2*\ba,-2*\ba) {};
  \coordinate (4) at (-\ba,-\ba) {};
  \coordinate (5) at (0,-2*\ba) {};
  \coordinate (6) at (\ba,-3*\ba) {};
  \coordinate (7) at (-2*\ba,-2*\ba) {};
  \coordinate (8) at (-\ba,-3*\ba) {};
  \coordinate (9) at (0,-4*\ba) {};

  \coordinate (21) at (0 + \ec, 0){};
  \coordinate (22) at (\ba + \ec,-\ba) {};
  \coordinate (23) at (2*\ba + \ec,-2*\ba) {};
  \coordinate (24) at (-\ba + \ec,-\ba) {};
  \coordinate (25) at (0 + \ec,-2*\ba) {};
  \coordinate (26) at (\ba + \ec,-3*\ba) {};
  \coordinate (27) at (-2*\ba + \ec,-2*\ba) {};
  \coordinate (28) at (-\ba + \ec,-3*\ba) {};
  \coordinate (29) at (0 + \ec,-4*\ba) {};

  \coordinate (r0) at (0.5*\ba + \ec,-0.5*\ba ) {};
  \coordinate (r1) at (1.5*\ba + \ec,-1.5*\ba ) {};
  \coordinate (r2) at (-.5*\ba + \ec,-3.5*\ba ) {};
  \coordinate (r3) at (-1.5*\ba + \ec,-2.5*\ba ) {};

  \coordinate (x) at (4,-3) {};
  \coordinate (y) at (0,1) {};



  \draw[-, line width=\wi mm] (1) -- (3);
  \draw[-, line width=\wi mm] (9) -- (3);
  \draw[-, line width=\wi mm] (1) -- (7);
  \draw[-, line width=\wi mm] (9) -- (7);

  \draw[-, line width=\wi mm] (21) -- (23);
  \draw[-, line width=\wi mm] (29) -- (23);
  \draw[-, line width=\wi mm] (21) -- (27);
  \draw[-, line width=\wi mm] (29) -- (27);

  \draw[-, lblue, line width=\wi mm] (1) -- (6);
 \draw[-, lblue, line width=\wi mm] (6) -- (9);
  \draw[-, lblue, line width=\wi mm] (9) -- (4);
  \draw[-, lblue, line width=\wi mm] (1) -- (4);


  \draw[-, lpink, line width=\wi mm] (r0) -- (r1);
  \draw[-, lpink, line width=\wi mm] (r2) -- (r1);
  \draw[-, lpink, line width=\wi mm] (r0) -- (r3);
  \draw[-, lpink, line width=\wi mm] (r2) -- (r3);

  \path[every node/.style={font=\sffamily\small}]
    (1) edge [draw=blue, bend left, line width=\wi mm] node [left] {} (6)
    (3) edge [draw=blue, bend left, line width=\wi mm] node[right] {} (9)
    (21) edge [draw=pink, bend left, line width=\wi mm] node [left] {} (r1)
    (23) edge [draw=pink, bend left, line width=\wi mm] node[right] {} (r2)

;
\end{tikzpicture}
\caption{Left: The polyhedron $\mathcal{Q}$ (black), $A \mathcal{Q}$ (light grey). The dark grey arrows represent  $ A (1 \quad 0)^\top = (0 \;\; -1)^\top$ and $ A (0 \quad 1)^\top = \left(\frac{1}{2} \quad -\frac{1}{2}\right)^\top$. Right: the same for matrix $B$.}
\label{f1}
\end{figure}

To make the construction easier to follow, we start with one node for each face (instead of one for each pair of opposite faces). The graph has therefore nine nodes: one for $\text{int}(\mathcal{Q})$, one for each vertex (the corners) and one for each facet (the sides of the square). The image by $A$ of vertex $F_1 = \begin{pmatrix} 1 & 0 \end{pmatrix}^\top$ is $\begin{pmatrix} \frac{1}{2} & -\frac{1}{2} \end{pmatrix}^\top$, which is in the face $$F_2 = \left\{x \; | \; x_1 - x_2 = 1, \;  x_1 + x_2 < 1, \;  - x_1 - x_2 < 1\right\}.$$
There is therefore an edge from the node representing $F_1$ to the node representing $F_2$, as depicted on the left of Figure \ref{f2}. By doing the same for each face, we find the entire graph for matrix $A$ and polyhedron $\mathcal{Q}$. 

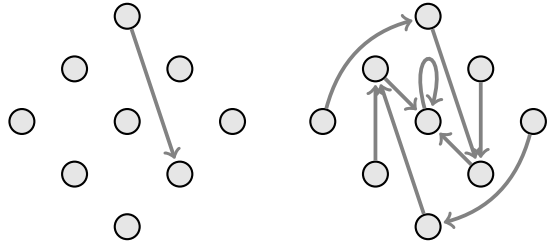
\begin{figure}[h!]
\centering

\usetikzlibrary{arrows}
\newcommand{\ec}{4}
\newcommand{\ba}{.7}
\newcommand{\wi}{.5}
\begin{tikzpicture}[->,shorten >=1pt,auto,node distance=2cm,
  thick,main node/.style={circle,fill=blue!20,draw,font=\sffamily\Large\bfseries}]


  \node[main node] (41) at (0 + 2*\ec, 0){};
  \node[main node] (42) at (\ba + 2*\ec,-\ba) {};
  \node[main node] (43) at (2*\ba + 2*\ec,-2*\ba) {};
  \node[main node] (44) at (-\ba + 2*\ec,-\ba) {};
  \node[main node] (45) at (0 + 2*\ec,-2*\ba) {};
  \node[main node] (46) at (\ba + 2*\ec,-3*\ba) {};
  \node[main node] (47) at (-2*\ba + 2*\ec,-2*\ba) {};
  \node[main node] (48) at (-\ba + 2*\ec,-3*\ba) {};
  \node[main node] (49) at (0 + 2*\ec,-4*\ba) {};

  \node[main node] (61) at (0 + 3*\ec, 0){};
  \node[main node] (62) at (\ba +3*\ec,-\ba) {};
  \node[main node] (63) at (2*\ba + 3*\ec,-2*\ba) {};
  \node[main node] (64) at (-\ba + 3*\ec,-\ba) {};
  \node[main node] (65) at (0 + 3*\ec,-2*\ba) {};
  \node[main node] (66) at (\ba + 3*\ec,-3*\ba) {};
  \node[main node] (67) at (-2*\ba + 3*\ec,-2*\ba) {};
  \node[main node] (68) at (-\ba + 3*\ec,-3*\ba) {};
  \node[main node] (69) at (0 + 3*\ec,-4*\ba) {};



  \path[every node/.style={font=\sffamily\small}]

    (41) edge [draw=blue, line width=\wi mm] node [left] {} (46)

    (61) edge [draw=blue, line width=\wi mm] node [left] {} (66)
    (63) edge [draw=blue, bend left, line width=\wi mm] node[right] {} (69)
    (62) edge [draw=blue, line width=\wi mm] node [left] {} (66)
    (64) edge [draw=blue, line width=\wi mm] node[right] {} (65)
    (65) edge [draw=blue, loop above, distance=9mm, line width=\wi mm] node [left] {} (65)
    (66) edge [draw=blue, line width=\wi mm] node[right] {} (65)
    (67) edge [draw=blue, bend left, line width=\wi mm] node[right] {} (61)
    (68) edge [draw=blue, line width=\wi mm] node[right] {} (64)
    (69) edge [draw=blue, line width=\wi mm] node[right] {} (64)


%

    
%
%

;
\end{tikzpicture}
\caption{Left: the nodes of the graph with the edge from $F_1$ to $F_2$. The node in the middle represents $\text{int}(\mathcal{Q})$, the nodes in the corners represent the vertices of the polyhedron and the other nodes represent the facets of the polyhedron. The edge is the one from $F_1$ to $F_2$.    Right: the graph with all edges for matrix $A$ (one edge from each node).}
\label{f2}
\end{figure}

We then add the edges corresponding to matrix $B$ (Figure \ref{f3}, left). 
The last step is to merge the nodes representing opposite faces and removing the edges that appear twice. We obtain the final graph of faces (Figure \ref{f3}, right). 

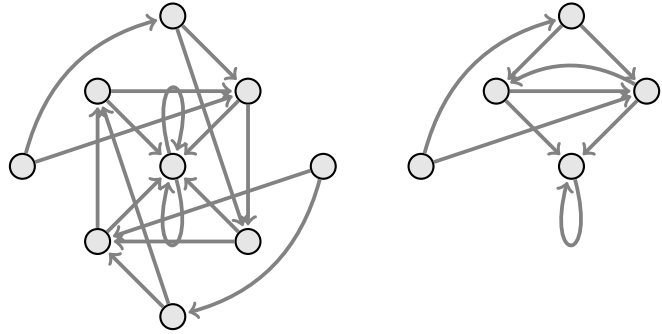
\begin{figure}[h!]
\centering

\usetikzlibrary{arrows}
\newcommand{\ec}{5.3}
\newcommand{\ba}{1}
\newcommand{\wi}{.5}
\begin{tikzpicture}[->,shorten >=1pt,auto,node distance=2cm,
  thick,main node/.style={circle,fill=blue!20,draw,font=\sffamily\Large\bfseries}]

  \node[main node] (41) at (0 + 2*\ec, 0){};
  \node[main node] (42) at (\ba + 2*\ec,-\ba) {};
  \node[main node] (43) at (2*\ba + 2*\ec,-2*\ba) {};
  \node[main node] (44) at (-\ba + 2*\ec,-\ba) {};
  \node[main node] (45) at (0 + 2*\ec,-2*\ba) {};
  \node[main node] (46) at (\ba + 2*\ec,-3*\ba) {};
  \node[main node] (47) at (-2*\ba + 2*\ec,-2*\ba) {};
  \node[main node] (48) at (-\ba + 2*\ec,-3*\ba) {};
  \node[main node] (49) at (0 + 2*\ec,-4*\ba) {};

  \node[main node] (61) at (0 + 3*\ec, 0){};
  \node[main node] (62) at (\ba +3*\ec,-\ba) {};
  \node[main node] (64) at (-\ba + 3*\ec,-\ba) {};
  \node[main node] (65) at (0 + 3*\ec,-2*\ba) {};
  \node[main node] (67) at (-2*\ba + 3*\ec,-2*\ba) {};




  \path[every node/.style={font=\sffamily\small}]

    (41) edge [draw=blue, line width=\wi mm] node [left] {} (46)
    (43) edge [draw=blue, bend left, line width=\wi mm] node[right] {} (49)
    (42) edge [draw=blue, line width=\wi mm] node [left] {} (46)
    (44) edge [draw=blue, line width=\wi mm] node[right] {} (45)
    (45) edge [draw=blue, loop above, distance=12mm, line width=\wi mm] node [left] {} (45)
    (46) edge [draw=blue, line width=\wi mm] node[right] {} (45)
    (47) edge [draw=blue, bend left, line width=\wi mm] node[right] {} (41)
    (48) edge [draw=blue, line width=\wi mm] node[right] {} (44)
    (49) edge [draw=blue, line width=\wi mm] node[right] {} (44)

    (61) edge [draw=blue, line width=\wi mm] node [left] {} (64)
    (64) edge [draw=blue, line width=\wi mm] node[right] {} (65)
    (67) edge [draw=blue, bend left, line width=\wi mm] node[right] {} (61)

    (62) edge [draw=blue, bend right, line width=\wi mm] node [left] {} (64)

    (41) edge [draw=pink, line width=\wi mm] node [left] {} (42)
    (43) edge [draw=pink, line width=\wi mm] node[right] {} (48)
    (42) edge [draw=pink, line width=\wi mm] node [left] {} (45)
    (44) edge [draw=pink, line width=\wi mm] node[right] {} (42)
    (46) edge [draw=pink, line width=\wi mm] node [left] {} (48)
    (47) edge [draw=pink, line width=\wi mm] node[right] {} (42)
    (48) edge [draw=pink, line width=\wi mm] node [left] {} (45)
    (49) edge [draw=pink, line width=\wi mm] node[right] {} (48)
    (45) edge [draw=pink, loop below, distance=12mm, line width=\wi mm] node [left] {} (45)

    (61) edge [draw=pink, line width=\wi mm] node [left] {} (62)
    (62) edge [draw=pink, line width=\wi mm] node [left] {} (65)
    (64) edge [draw=pink, line width=\wi mm] node[right] {} (62)
    (67) edge [draw=pink, line width=\wi mm] node[right] {} (62)
    (65) edge [draw=pink, loop below, distance=12mm, line width=\wi mm] node [left] {} (65)

%

    
%
%

;
\end{tikzpicture}
\caption{Left : the graph for matrices $A$ and $B$.  Right : the graph of faces as defined in Definition \ref{def_gr}.}
\label{f3}
\end{figure}


\end{ex} \color{black}

\vspace{3cm}

\begin{thmpri} Let $S$ be a set of matrices satisfying $A\mathbf{1} = \mathbf{1}$ and Assumption 1. The answer to Problem 1 is positive if and only if the 
 self-loop of node 1 is the only cycle in the graph of faces.
\begin{proof}
We have seen in Proposition \ref{nec_suf} that the answer to Problem 1 is negative if and only if
\begin{equation*} \exists \; \textnormal{proper face } F, \; \sigma \in \Sigma, \; k \leq N, \; A_{\sigma(k-1)} \dots A_{\sigma(0)}F \subseteq \pm F.  \end{equation*}
This condition is equivalent to the nonexistence of a cycle other than the self-loop of node 1.
\end{proof}
\label{main_thm}
\end{thmpri}

\begin{thmbis}{main_thm} Let $S$ be a set of matrices satisfying $A\mathbf{1} = \mathbf{1}$ and Assumption 1. The answer to Problem 2 is positive if and only if there is a path from any node to the node 1.
\begin{proof} We use Proposition \ref{nec_suf2} and the equivalence between
\begin{equation*}\forall x \in \mathcal{P}, \; \exists \sigma \in \Sigma, \; A_{\sigma(N-1)} \dots A_{\sigma(0)}x \subseteq \textnormal{int}(\mathcal{P})\end{equation*} and the fact that from any node, there is a path of length $N$ leading to node~1. 
\end{proof}
\label{main_thmbis}
\end{thmbis}

\section{Computational aspects}
\label{Complexity}

Now that we have necessary and sufficient conditions (\ref{main_thm}, \ref{main_thmbis}) for Problems 1 and 2, we estimate the algorithmic complexity of evaluating these conditions. 

To construct the graph of faces, we need two basic operations: to compute in which face a point is, and to find a point in a given face.
In Lemma \ref{v_repr}, we have seen that there is a one to one correspondence between the proper open faces of $\mathcal{P}$ and the elements of $V$. These elements will be used to represent the proper faces. 
From this representation, it is computationally easy (in $O(n)$) to determine in which open face a point $x \in \mathcal{P}$ is: 
\begin{itemize}
\item Compute $\|x\|_\mathcal{P} = \max_i x_i - \min_i x_i$ to determine if $x \in \text{int}(\mathcal{P})$
\item if not, by Lemma \ref{v_repr}, $v = \textnormal{round}(x + (1 - \max_i x_i ) \mathbf{1})$ gives the face in which $x$ is.
\end{itemize} Finding a point in an open face can be done by just taking $v$ itself.

We are now able to prove our complexity result.

\begin{thm} Problems 1 and 2 can be decided in $O(3^n m n^2)$ operations.
\begin{proof} \textit{Construction of the graph of faces:} The graph has $N = \frac{1}{2} (3^n - 2^{n+1} + 1)$ nodes. Each node has at most $m$ outgoing edges (at most $N m$ in total), corresponding to the $m$ transition matrices in $S$. 

To compute the edge starting from node $i$ (representing face $F$) and corresponding to transition matrix $A$, we need to find the face $G$ such that $$AF \subseteq G.$$ By Lemma \ref{key}, we know that $G$ is the face containing $Ax$ where $x$ is any point in $F$. 
Finding $G$ can be done in $O(n^2)$ operations : take a point $x$ in $F$ (in $O(n)$), compute $Ax$ (in $O(n^2)$), and find the face 
 in which $Ax$ is (in $O(n)$). Therefore the complexity of constructing the graph of faces is $O(3^n m n^2)$.

\textit{Decision problems on the graph:} Once the graph is constructed, Problem 1, which is equivalent to the existence of cycles in the graph (see Theorem \ref{main_thm}), can be decided using a topological sorting algorithm which has a complexity of $O(|E| + |V|) = O(3^n m) $ \cite{topo}. Problem 2, which is equivalent to the connectivity of the graph (see Theorem \ref{main_thmbis}), can be decided using a search algorithm which has the same complexity. The total complexity is therefore dominated by the complexity of the construction of the graph : $O(3^n m n^2)$.
\end{proof}
\end{thm}

From \cite{BO, MTNS}, the problem was known to be decidable in  $O(m^{3^n} n^\omega)$ operations where $O(n^\omega)$ is the complexity of multiplying two $n \times n$ matrices. We have now an algorithm with singly exponential complexity.

\section{Sets of two stochastic undirected matrices}
\label{new_part}
In this section, we study the effect of reciprocity on the asymptotic stability problem (Problem 1). We restrict our attention to sets of two stochastic undirected matrices.  A nonnegative matrix is said to be undirected if $$a_{ij} > 0 \Leftrightarrow a_{ji} > 0.$$ A stochastic matrix is a nonnegative matrix satisfying $A \mathbf{1} = \mathbf{1}$. 
It is known that reciprocity plays an important role in the convergence of consensus systems \cite{cut}. 
It has been proven that Problem 1 is NP-hard for sets of three undirected matrices  and for sets of two matrices in general \cite{BO}. The authors have left open the case of sets of two matrices. We prove that it can be solved in polynomial time. 

It is worth noticing that the product of two stochastic matrices is a stochastic matrix, and that stochastic matrices satisfy the relations:
$$\max_i (Ax)_i \leq \max_i x_i$$
$$\min_i (Ax)_i \geq \min_i x_i$$
and therefore they satisfy Assumption 1.

The next lemma presents a simple yet crucial observation about undirected stochastic matrices.
\begin{lem} Let $A$ be an undirected stochastic matrix. Then $A^2$ has a positive diagonal.
\label{pos_diag}
\end{lem}

The following lemma shows the effect of a transition matrix with a positive diagonal. The multiplication by a stochastic matrix with positive diagonal cannot activate any of the facet constraints of $\mathcal{P}$.

\begin{lem}
Let $x \in \mathcal{P}$, let $\bar{F}$ be a (closed) face of $\mathcal{P}$, and let $ B$ be a stochastic matrix that has a positive diagonal, 
$$Bx \in \bar{F} \Rightarrow x \in \bar{F}.$$
Furthermore, with $A$ and $C$ any stochastic matrices, we have $$ABCx \in \bar{F} \Rightarrow ACx \in \bar{F}.$$
\begin{proof} 

We first prove that 
\begin{equation} x_i - x_j < 2 \Rightarrow (Bx)_i - (Bx)_j < 2 \label{step2} \end{equation} meaning that if the facet constraint $x_i - x_j \leq 2$ is not active for $x$, it cannot be active for $Bx.$

Using $\sum_k B_{ik} = 1$ (because the matrix is stochastic) and $a$ any number satisfying $0 < a \leq \min(B_{ii}, B_{jj})$, we obtain
\begin{equation} \begin{aligned} (Bx)_i & = \sum_k B_{ik} x_k  \\
& \leq B_{ii} x_i + \sum_{k \neq i} B_{ik} \max_l x_l  \\
& = B_{ii} x_i + (1 - B_{ii}) \max_l x_l \\
& \leq a x_i + (1 - a) \max_l x_l
\end{aligned} \label{Bx} \end{equation}
where the last inequality comes from $a\leq B_{ii}$ and $x_i \leq \max_l x_l$.

Similarly, we have 
\begin{equation}(Bx)_j \geq a x_j + (1 - a) \min_l x_l \label{similarly} \end{equation}
and combining (\ref{Bx}), (\ref{similarly}) and the hypotheses $x_i - x_j < 2$ and $\max_l x_l - \min_l x_l \leq 2$ (because $x \in \mathcal{P}$), we obtain
$$(Bx)_i - (Bx)_j \leq a (x_i-x_j) + (1 - a) (\max_l x_l - \min_l x_l) < 2.$$

Since a face is an intersection of facets, if $\bar{F}$ is a face such that $Bx \in \bar{F}$, then using the contrapositive of (\ref{step2}) for all the facets yields $x \in \bar{F}$ and the first part of the lemma is proved.

It is clear that $Cx$ is a point, so that applying the first part of the lemma to $Cx$ yields

$$BCx \in \bar{F} \Rightarrow Cx \in \bar{F}.$$

Now if $ABCx \in \bar{F}$, there is a proper closed face $\bar{G}$ such that $BCx \in \bar{G}$ and therefore $Cx \in \bar{G}$. By Lemma \ref{key}, the image of $G = \text{int}(\bar{G})$ by $A$ is a subset of $F = \text{int}(\bar{F})$ and therefore $A\bar{G} \subseteq \bar{F}$ and $ACx \in \bar{F}.$

\end{proof}
\label{closure}
\end{lem}

\begin{prop}
Let $S = \{A_1,  A_2\}$ be a set of two stochastic undirected matrices. The answer to Problem 1 is positive if and only if System \ref{sys} converges for any initial condition for the sequences \begin{equation*} \begin{aligned}
\sigma_1 & = 1, 1, \dots \label{seq1} \\
\sigma_2 & = 2, 2, \dots \\
\sigma_3 & = 1, 2, 1, 2 \dots 
\end{aligned}
\end{equation*}
\begin{proof}
It is clear that if the system does not converge for one of the three sequences, then System \ref{sys} does not converge for any sequence and the answer to Problem 1 is negative. Therefore, we only prove that the convergence for the three sequences is sufficient for a positive answer.

By Lemma \ref{pos_diag}, $A_1^2$ and $A_2^2$ have positive diagonal. 
  
With these claims in mind, we can prove the proposition.
Suppose that the answer to Problem 1 is negative.
By Proposition \ref{nec_suf}, there is a face $F$ and a product $A_{\sigma(k-1)} \dots A_{\sigma(0)}$ of finite length $k $ such that $$A_{\sigma(k-1)} \dots A_{\sigma(0)}F \subseteq F.$$ 
Taking the square of the product provides a product of even length $2k$ having this property:
$$(A_{\sigma(k-1)} \dots A_{\sigma(0)})^2F \subseteq F \subseteq \text{cl}(F).$$
Let us now take the shortest sequence $\sigma^*$ of even length $k^*$ satisfying
\begin{equation} A_{\sigma^*(k^{*}-1)} \dots A_{\sigma^*(0)}F \subseteq \text{cl}(F). \label{magic_prop} \end{equation} This product has a length of at least 4. Otherwise it would be $A_1 A_2$, $A_2 A_1$, $A_1^2$ or $A_2^2$, so that the system would not converge for one of the sequence $\sigma_1,\sigma_2,\sigma_3$, contradicting the hypothesis.
Suppose that this product contains the product $A_i^2$:
$$A_{\sigma^*(k^{*}-1)} \dots A_{\sigma^*(l)} A_i A_i A_{\sigma^*(l-3)} \dots A_{\sigma^*(0)} F \subseteq \text{cl}(F).$$
The set $\text{cl}(F)$ is a closed face and we can use Lemma \ref{closure} to obtain 
$$A_{\sigma^*(k^{*}-1)} \dots A_{\sigma^*(l)}  A_{\sigma^*(l-3)} \dots A_{\sigma^*(0)} F \subseteq \text{cl}(F)$$ so that there is a shorter product of even non-zero length satisfying (\ref{magic_prop}) contradicting the fact that we took the shortest one. 
In turn, $A_{\sigma^*(k^{*}-1)} \dots A_{\sigma^*(0)}$ does not contain $A_1^2$ nor $A_2^2$: $$A_{\sigma^*(k^{*}-1)} \dots A_{\sigma^*(0)} = (A_1 A_2)^{k/2} \text{ or } (A_2 A_1)^{k/2},$$ contradicting now the hypothesis of the proposition.

\end{proof}
\end{prop}

We can determine if all trajectories converge by verifying the convergence of powers of individual matrices ($A_1$, $A_2$ and $A_1 A_2$). This can be done by computing the second eigenvalue of each matrix, hence the next corollary.
\begin{cor}
Problem 1 can be decided in polynomial time for sets of two stochastic undirected matrices.
\end{cor}

\section{Conclusion}

The goal of this paper was to investigate the complexity of determining if the convergence of a consensus system is guaranteed and of determining if the convergence is possible. We have obtained a geometric characterization allowing for singly exponential algorithms for both problems. By doing so, we have improved the known complexity of Problem 1, which was doubly exponential. This first problem is also known to be NP-hard so there was little hope to obtain a much better complexity.

Another case in which the complexity of Problem 1 was open is for sets of two undirected matrices. 
We proved the existence of a polynomial-time algorithm for this case.

To obtain these results, we have introduced the graph of faces but the possibilities offered by  this new object remain largely unexplored. In particular, simulations suggest that graph of faces have particular structures that could be used to solve other problems or to find faster algorithms.

Consensus systems with stochastic matrices have an invariant polyhedron making them naturally suited for the analysis that we have developed.
We would like to mention however that this reasoning can apply for any discrete time linear switched system that admits a common invariant polyhedron. The finiteness would still hold and in most cases so would the singly exponential complexity.  The exact complexity may be different. Indeed, one of the building block of the method is to determine in which face a point is. We can do it here in $O(n)$ operations because of the representation of the polyhedron given by Lemma \ref{v_repr}. This compact representation is possible for this particular polyhedron but not necessarily for all of them.

\end{document}